\documentclass[%
 reprint,
superscriptaddress,
 amsmath,amssymb,
 aps,showkeys,https://www.overleaf.com/project/608bd335d666745eb22d6f68
]{revtex4-2}

\usepackage{xcolor}

\usepackage[ruled]{algorithm2e}
\usepackage{siunitx}
\usepackage{graphicx}
\usepackage{dcolumn}
\usepackage{bm}
\usepackage{hyperref}
\usepackage{multirow}
\usepackage{mathtools}
\usepackage{diagbox}
\usepackage{tikz}
\usetikzlibrary{shapes, arrows}
\usepackage[bottom]{footmisc}
\usepackage{xcolor}
\hypersetup{
    colorlinks,
    linkcolor={blue!60!black},
    citecolor={blue!60!black},
    urlcolor={blue!60!black}
}
\usepackage{subfig}

\DeclareMathOperator*{\argmax}{\arg\!\max}

\tikzstyle{startstop} = [rectangle, rounded corners, 
minimum width=3cm, 
minimum height=1cm,
text centered, 
draw=black, 
fill=red!30]

\tikzstyle{io} = [trapezium, 
trapezium stretches=true, 
trapezium left angle=70, 
trapezium right angle=110, 
minimum width=3cm, 
minimum height=1cm, text centered, 
draw=black, fill=blue!30]

\tikzstyle{process} = [rectangle, 
minimum width=3cm, 
minimum height=1cm, 
text centered, 
text width=3cm, 
draw=black, 
fill=orange!30]

\tikzstyle{decision} = [diamond, 
minimum width=3cm, 
minimum height=1cm, 
text centered, 
draw=black, 
fill=green!30]
\tikzstyle{arrow} = [thick,->,>=stealth]
\tikzset{
    rectangle connector/.style={
        connector,
        to path={(\tikztostart) -- ++(#1,0pt) \tikztonodes |- (\tikztotarget) },
        pos=1.5
    },
    rectangle connector/.default=-2cm,
    straight connector/.style={
        connector,
        to path=--(\tikztotarget) \tikztonodes
    }}

\newcommand{\esa}{European Space Agency (ESA), European Space Research and Technology Centre (ESTEC), Keplerlaan 1, 2201 AZ Noordwijk, the Netherlands}

\begin{document}

\title{
Searching for extreme mass ratio inspirals in LISA:\\ from identification to parameter estimation
}

\author{Stefan H. Strub}
 \email{stefan.strub@eaps.ethz.ch}
\affiliation{%
Institute of Geophysics, ETH Zurich\\ Sonneggstrasse 5, 8092 Zurich, Switzerland
}

\author{Lorenzo Speri}\affiliation{\esa}

\author{Domenico Giardini}
\affiliation{Institute of Geophysics, ETH Zurich\\ Sonneggstrasse 5, 8092 Zurich, Switzerland
}


%

\begin{abstract}
The Laser Interferometer Space Antenna (LISA) is a planned space-based observatory designed to detect gravitational waves (GWs) within the millihertz frequency range. LISA is anticipated to observe the inspiral of compact objects into black holes at the centers of galaxies, so called extreme-mass-ratio inspirals (EMRIs). However, the extraction of these long-lived complex signals is challenging due to the large size and multimodality of the search space. In this study, we introduce a new search strategy that allows us to find EMRI signals in noisy data from wide priors all the way to performing parameter estimation. This work is an important step in understanding how to extract EMRIs from future LISA data.
\end{abstract}

\keywords{Gravitational Waves, LISA, Extreme-Mass-Ratio Inspirals, EMRI, LISA Data Challenge, Global Fit}
\maketitle

\section{Introduction}

Extreme Mass Ratio Inspirals (EMRIs) are one of the key gravitational wave sources that the Laser Interferometer Space Antenna (LISA) is expected to detect \cite{Babak:2017tow,lisaredbook}. These signals originate from the inspiral of a stellar-origin compact object into a massive black hole (SMBH) typically residing at the center of a galaxy.
EMRI systems are characterized by extremely small mass ratios
$\mu/M\sim 10^{-4}-10^{-6}$, where $\mu\sim1-100M_\odot$ is the mass of the orbiting secondary and $M\sim10^{5}-10^{7}M_\odot$ is the mass of the primary massive black hole. 
The formation details depend on the precise formation mechanism \cite{Amaro-Seoane:2012lgq,Berry:2019wgg,Gair:2004iv}, but it is anticipated that most EMRIs will have non-negligible eccentricities at the plunge \cite{Babak:2017tow}.
Eccentric orbits and small mass ratios lead to gravitational wave signals characterized by a rich harmonic content and tens of thousands of cycles in band. This makes EMRIs unique sources for extracting information about the nature of gravity \cite{Speri:2024qak,Barack:2006pq,Sopuerta:2009iy,Speri:2024qak} and the surroundings of SMBHs \cite{Gair:2010yu,Speri:2022upm,Cole:2022yzw}.

There are several challenges associated with realizing this rich scientific potential. In this work, we focus on the challenge of searching and identifying EMRIs.
Unlike ground-based gravitational wave detectors, which can densely sample the parameter space using precomputed waveform banks, fully coherent template bank methods become impractical for EMRIs \cite{Gair:2004iv}. The posterior distribution characterizing EMRI signals occupies only a tiny fraction of this space, making it like searching for a needle in a cosmic haystack. As a result, conventional grid-based approaches are infeasible, leading to the adoption of semi-coherent searches and stochastic template placing.

Even with these techniques, identifying EMRI signals remains a complex task. The challenge arises from the presence of numerous local maxima within the search landscape. These local maxima are due to non-local parameter degeneracies in the signal space \cite{Chua:2021aah}. In simpler terms, many combinations of parameters can produce waveforms that closely match the data, creating these local peaks in the optimization landscape. However, these local maxima are different from the true global maxima, which are located at entirely different parameter configurations and are often separated by vast distances in the parameter space. Distinguishing true EMRI signals from these local peaks is a complex puzzle.

The Mock LISA Data Challenges have demonstrated the difficulties associated with the detection of EMRI signals \cite{babak2010mock}. Previous works have adopted several strategies to address these two problems. Stochastic searches from \cite{Babak:2009ua,Cornish:2008zd} got stuck on local maxima and tried to use such local maxima as information to move to the global maximum. An alternative approach was proposed in \cite{Chua:2022ssg}, where the use of a veto likelihood would remove the local maxima and allow only the global maxima to remain.
Other approaches used phenomenological waveforms, where the initial stage is to identify harmonics and then search for possible EMRI parameters \cite{Wang:2012xh}. 
Ref.~\cite{Badger:2024rld} introduced a sparse dictionary learning algorithm to reconstruct year-long EMRI waveforms. Ref.~\cite{Yun:2023vwa} developed a convolutional neural network for signal-to-noise ratios between 50 and 100, facilitating rapid detection and parameter estimation of EMRIs. Ref.~\cite{Zhang:2022xuq} demonstrated the applicability of a convolutional neural network in the detection of EMRI signals. Ref.~\cite{Ye:2023lok} combined template searches similar to \cite{Babak:2009ua,Cornish:2008zd} with phenomenological waveform seachers \cite{Wang:2012xh}. However, they manage to constrain the parameters to the region where most of the secondaries lie around the primary peak.

In this work, we make a step forward with respect to \cite{Ye:2023lok} by reaching the primary peak and performing parameter estimation, simulating a full search and parameter estimation pipeline for a single EMRI in stationary Gaussian noise.
We introduce a novel search statistic and employ it to achieve an accurate identification of EMRI signals, leading to the successful extraction of such signals. We define a successful signal extraction as a case where the difference between the recovered signal and the original GW signal in the data is significantly smaller in magnitude than the GW signal itself. This ensures that, after subtracting the recovered signal from the data, the residual can be effectively utilized to search for other GW signals. Furthermore, we performed parameter estimation using the parameters of the recovered signal, demonstrating the efficacy of our method in extracting EMRI signals and obtaining their posterior distributions.


\section{Methods}
\subsection{LISA Data Analysis}\label{sec:lisa_da}
The basic observables of LISA are the Time Delay Interferometry variables. These variables are constructed from six Doppler measurements in order to suppress laser frequency noise. In this work, we adopt the pseudo-orthogonal combinations $A$ and $E$ and exclude the channel $T$, which is suppressed below the transfer frequency $f_* = 1/(2 \pi L) \approx \SI{19.1}{mHz}$ \cite{prince2002lisa, vallisneri2005synthetic, littenberg2020global}.

We treat channels $A$ and $E$ as outputs from independent instruments. We assume that each channel is a time series, denoted as $d(t)$, and consists of the superposition of multiple gravitational wave (GW) signals and instrumental noise, $n(t)$. Here, we consider the case of a single EMRI source $s(t, \theta)$ present in the data:
\begin{equation}
    d(t) = s(t,\theta) + n(t).
\end{equation}
For simplicity, we omit explicit time dependence in subsequent expressions, considering $d$, $s(\theta)$, and $n$ as time series.

To simulate the signal component $s(\theta)$ of the LISA data, we use the FastEMRIWaveform package $\texttt{few}$ \cite{Chua:2020stf, michael_l_katz_2023_8190418, Chua:2018woh, Fujita:2020zxe, Stein:2019buj, Chua:2015mua, Chua:2017ujo, Barack:2003fp,speri2024fast} to obtain the gravitational wave signal in the solar system barycenter and the fast lisa response package \cite{chua_2020_3981654} to include the LISA response and produce the TDI outputs, and therefore obtain the signal $s(\theta)$. We consider EMRI waveforms of non-spinning black holes, which are described by 11 parameters:

\begin{equation}
\theta = \{ M, \mu, p_0, e_0, D_L, \theta_S, \phi_S, \Phi_{\varphi0}, \Phi_{r0}, \theta_K, \phi_K \}.
\end{equation}

Here, $M$ represents the mass of the central massive black hole, while $\mu$ denotes the mass of the secondary object. Orbital elements are characterized by the semi-latus rectum \( p_0 \) and eccentricity \( e_0 \). The source’s sky location in the Solar System barycenter frame is given by the angles \( \theta_S \) and \( \phi_S \), while the orientation of the binary’s orbital angular momentum is specified by \( \theta_K \) and \( \phi_K \). The luminosity distance is denoted by \( D_L \). The initial orbital phases are \( \Phi_{\varphi0} \) and \( \Phi_{r0} \).

\begin{table}[!ht]
\caption{Parameters of the injected EMRI.}
\begin{ruledtabular}
\begin{tabular}{ccc}
           Parameter &  Injected Value  \\ \hline
  $M$ ($\textup{M}_\odot$) & $10^6$\\
  $\mu$ ($\textup{M}_\odot$) &  $10$\\
  $e_0$ &  $0.2$\\
  $t_{p}$ (yr) &  $0.44$\\
  $p_{0}$ &  $8.5072$ \\
  $D_L$ (Gpc) & 0.85 \\
    $ \theta_S$  &     $3\pi / 4$\\
  $\phi_S$&      $3\pi / 4$\\
        $\Phi_{\varphi0}$&          $1$\\
        $\Phi_{r0}$&          $3$\\
    $\theta_K$  &     $3\pi / 4$\\
  $\phi_K$&      $3\pi / 4$\\
  $\rho_\textrm{optimal}$& $56$\\
\end{tabular}
\label{tab:injection}
\end{ruledtabular}
\end{table}

The simulated LISA data consist of an EMRI signal with the parameter values listed in Table \ref{tab:injection}. The intrinsic parameters of the EMRI are consistent with those used in \cite{ye2024identification}. The time to plunge $t_p$ is set to 0.44 years. 

The injection noise $n$ is modeled using \texttt{LISAtools} \cite{michael_katz_2024_10930980}, with the $\textit{FittedHyperbolicTangentGalacticForeground}$ model used to estimate the galactic noise over a six-month period.
Figure \ref{fig:data} presents the spectrogram of the EMRI signal injected into the TDI A channel alongside the noisy data for the input and the recovered EMRI signal.
Additionally, Figure~\ref{fig:noise} presents the noisy data, the EMRI signal, and the corresponding power spectral density (PSD) curve in the frequency domain, illustrating the reduction in noise around \SI{4}{mHz}.

\begin{figure}[!ht]
    \includegraphics[width=1.1\columnwidth]{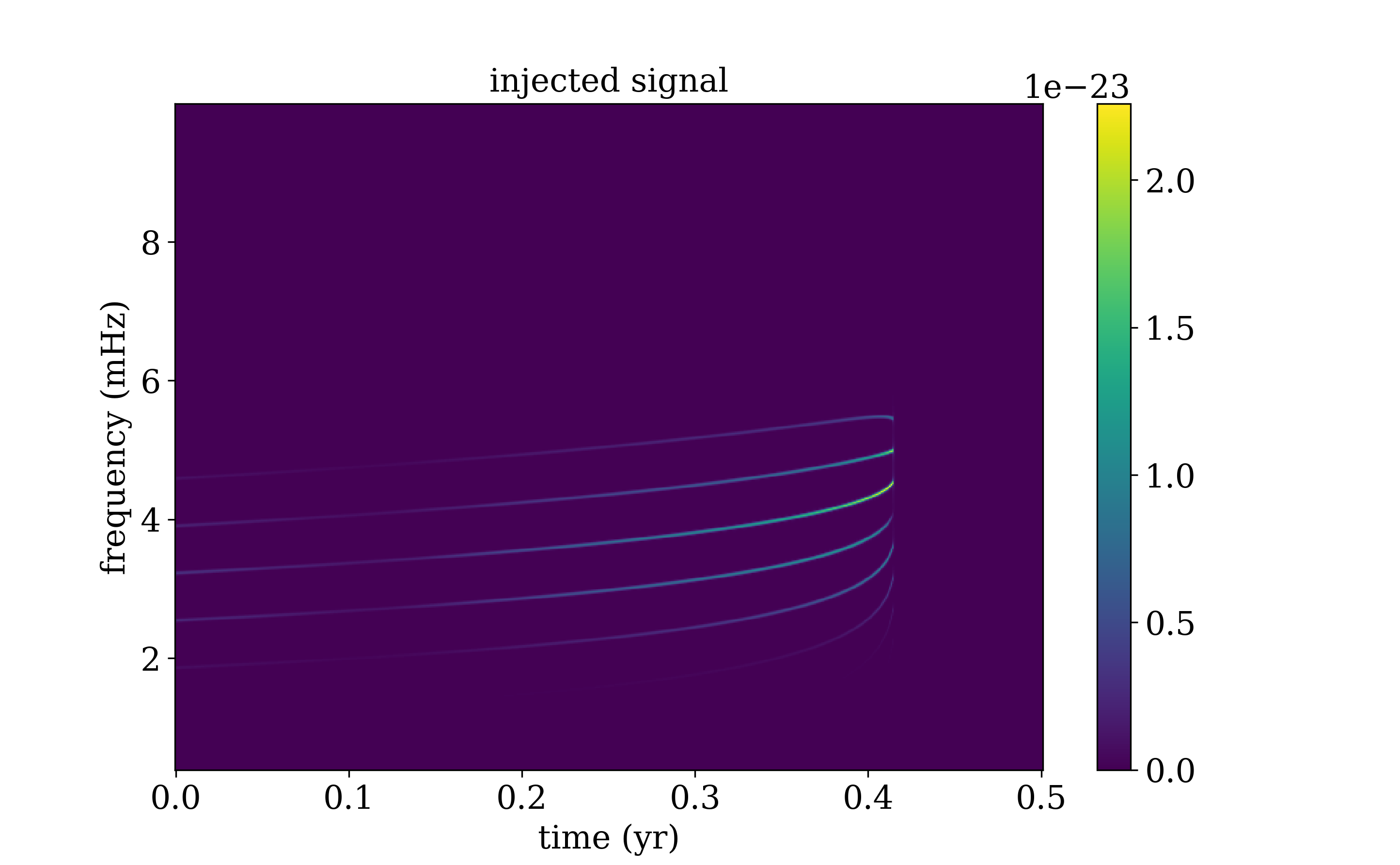}
    \includegraphics[width=1.1\columnwidth]{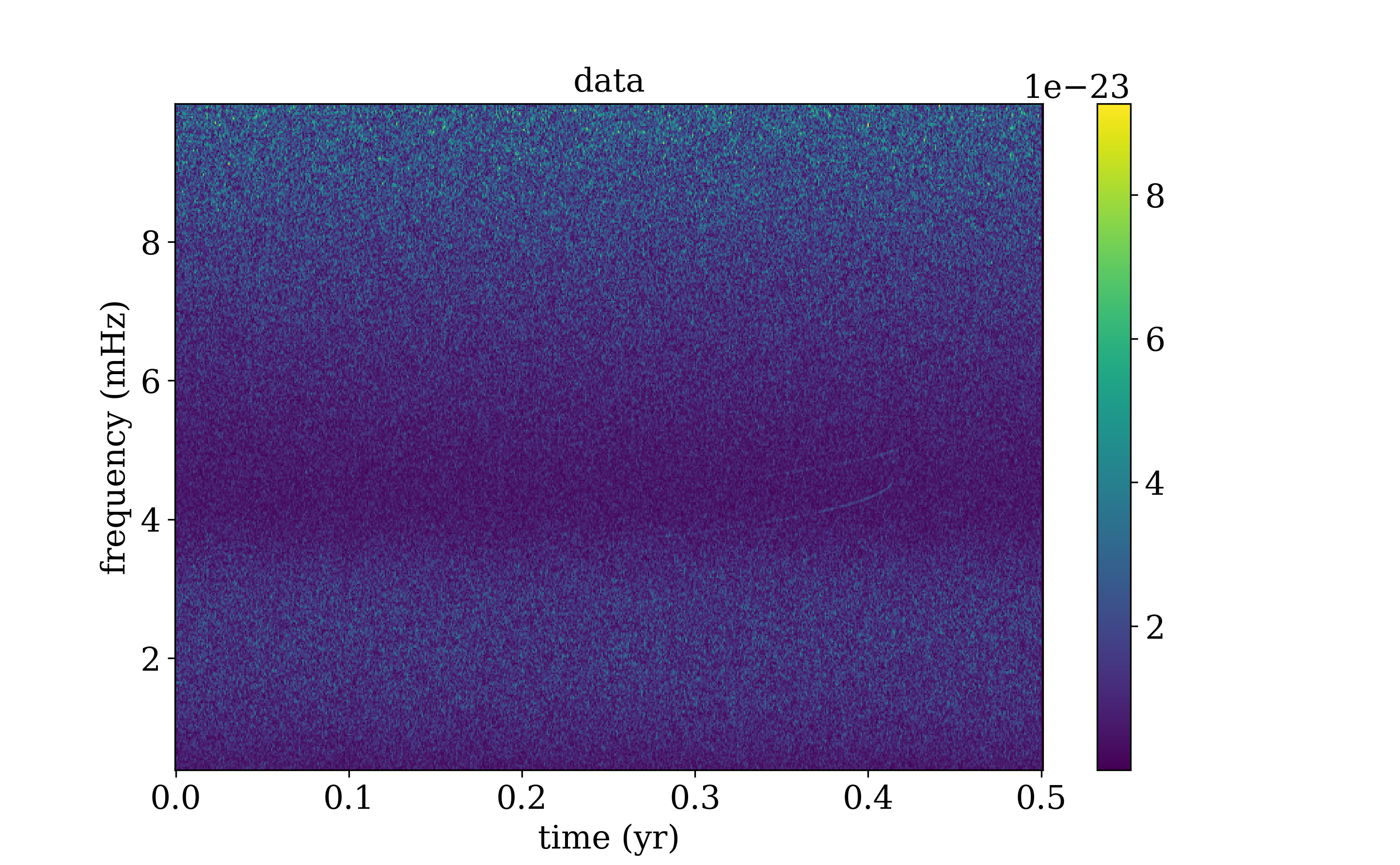}
    \includegraphics[width=1.1\columnwidth]{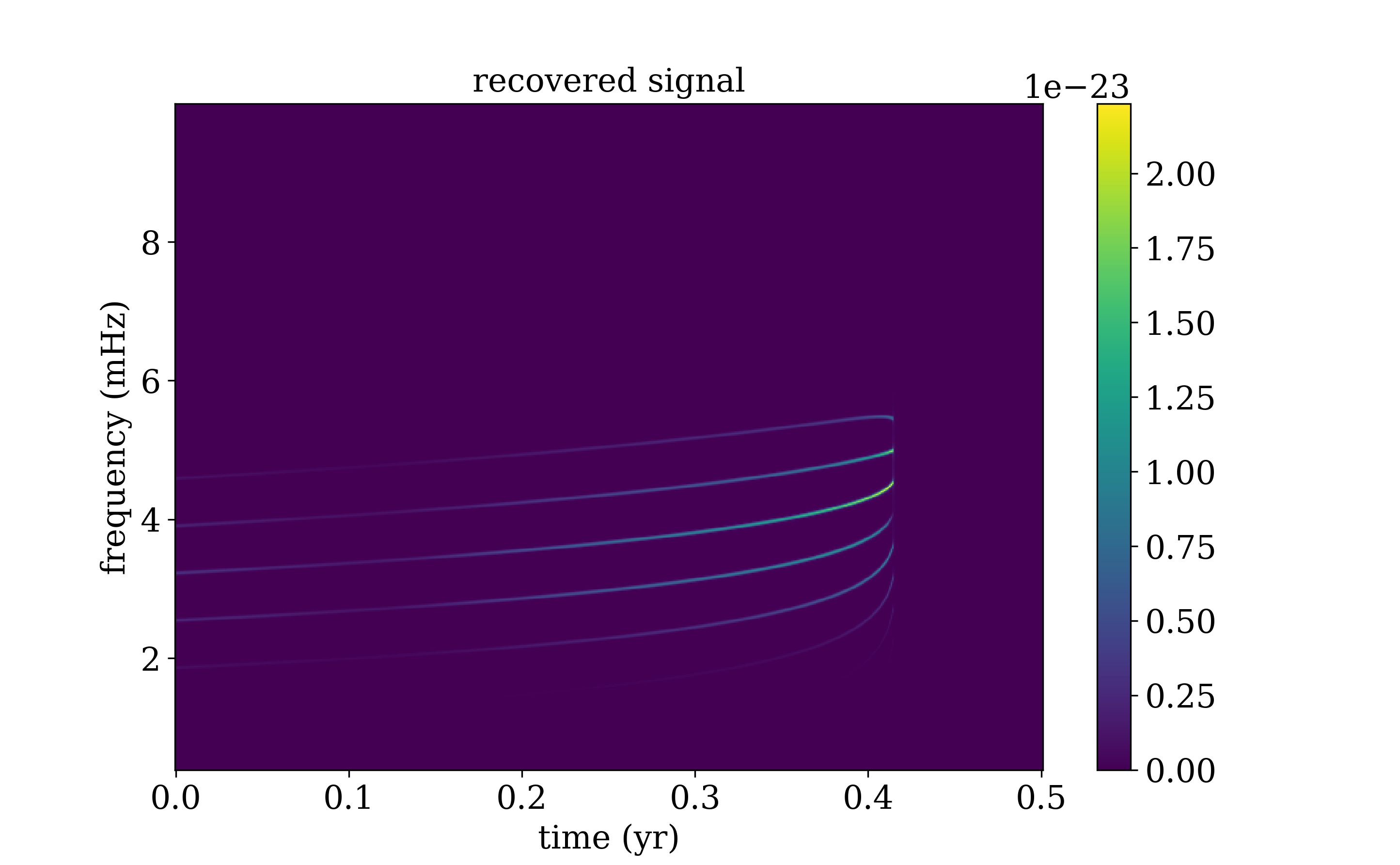}
\caption{Spectrogram of the TDI A channel of the noisy data, the injected and recovered signal. 
}
\label{fig:data}
\label{fig:data}
\end{figure}

\subsection{Likelihood and signal-to-noise ratio}
The likelihood function $p(d | \theta)$ quantifies the probability of observing data $d$ given the source parameters $\theta$. Under the assumption of stationary Gaussian noise, the likelihood for one channel can be obtained with
\begin{equation}
    \log p(d | \theta) \propto -\frac{1}{2} \langle d - s(\theta), d - s(\theta) \rangle,
\end{equation}
where the inner product of two signals $x(t)$ and $y(t)$ in the frequency domain is defined as:

\begin{equation}
\label{eq:scalar}
    \langle x, y \rangle = 4 \mathcal{R} \int_0^\infty \frac{\tilde{x}(f) \tilde{y}^*(f)}{S_n(f)} \ df,
\end{equation}

where $\tilde{x}(f)$ is the Fourier transform of $x(t)$, and $S_n(f)$ is the one-sided power spectral density of the noise \cite{michael_katz_2024_10930980}.







A proxy to find the best-fit parameters $\theta'$ given a dataset is to maximize the matched signal-to-noise ratio (SNR), given by:
\begin{equation}
\label{eq:SNR}
    \rho = \frac{\langle d, s(\theta') \rangle}{\sqrt{\langle s(\theta'), s(\theta') \rangle}},
\end{equation}
which is independent of the luminosity distance $D_L$. The optimal estimate for $D_L$ maximizing the likelihood is:
\begin{equation}
\label{eq:maxD}
    D_L = \frac{\langle s(\theta'), s(\theta') \rangle}{\langle d, s(\theta') \rangle},
\end{equation}

where $\theta' = \theta \setminus \{ D_L \}$. The optimal SNR is defined as:

\begin{equation}
    \rho_{\text{optimal}} \coloneqq \sqrt{\langle s(\theta), s(\theta) \rangle},
\end{equation}

which differs from the detected SNR due to noise fluctuations.

\subsection{Matching spectrograms}
\begin{figure*}[!ht]
    \centering
    \subfloat{\includegraphics[width=0.5\textwidth]{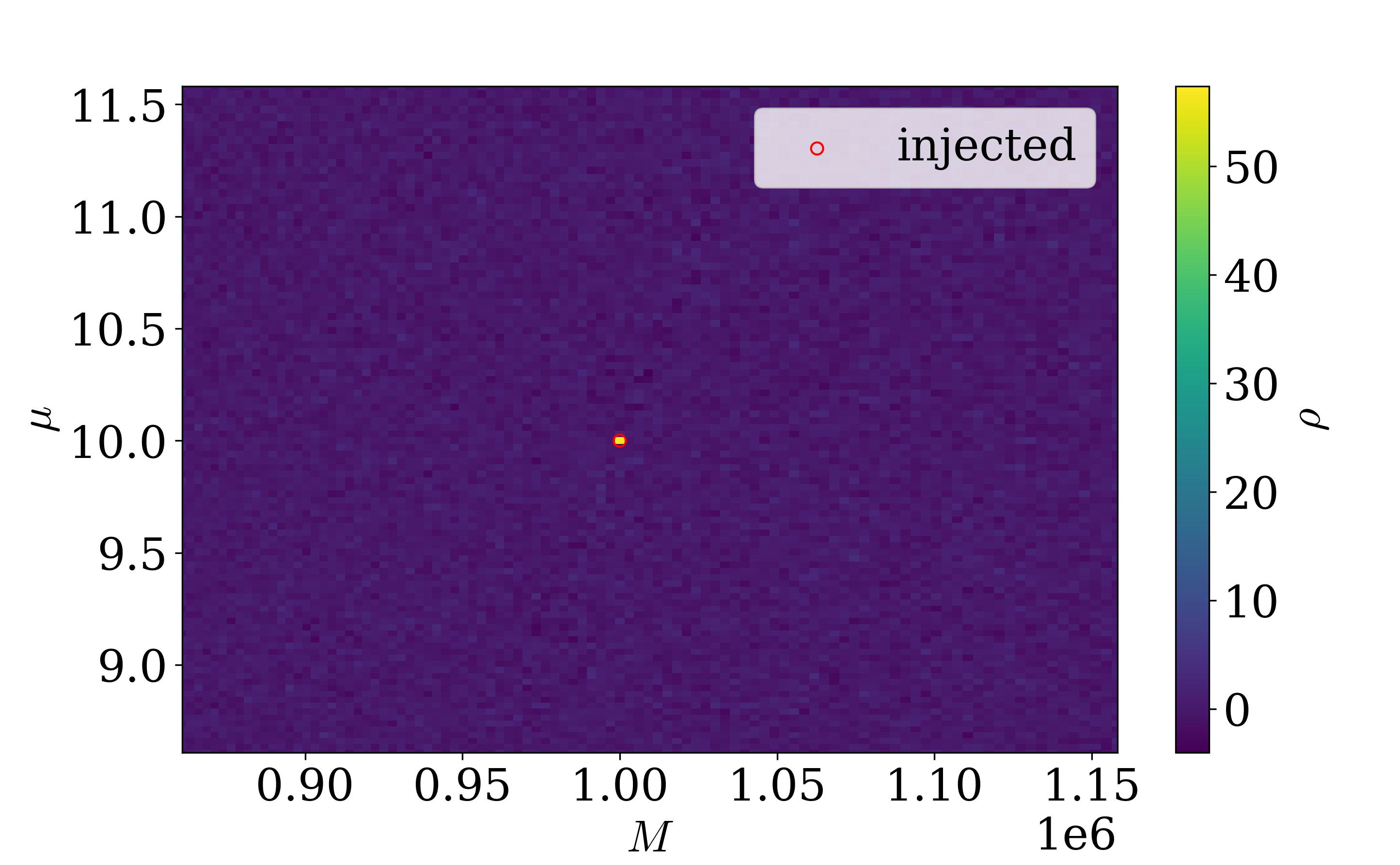}}
    \subfloat{\includegraphics[width=0.5\textwidth]{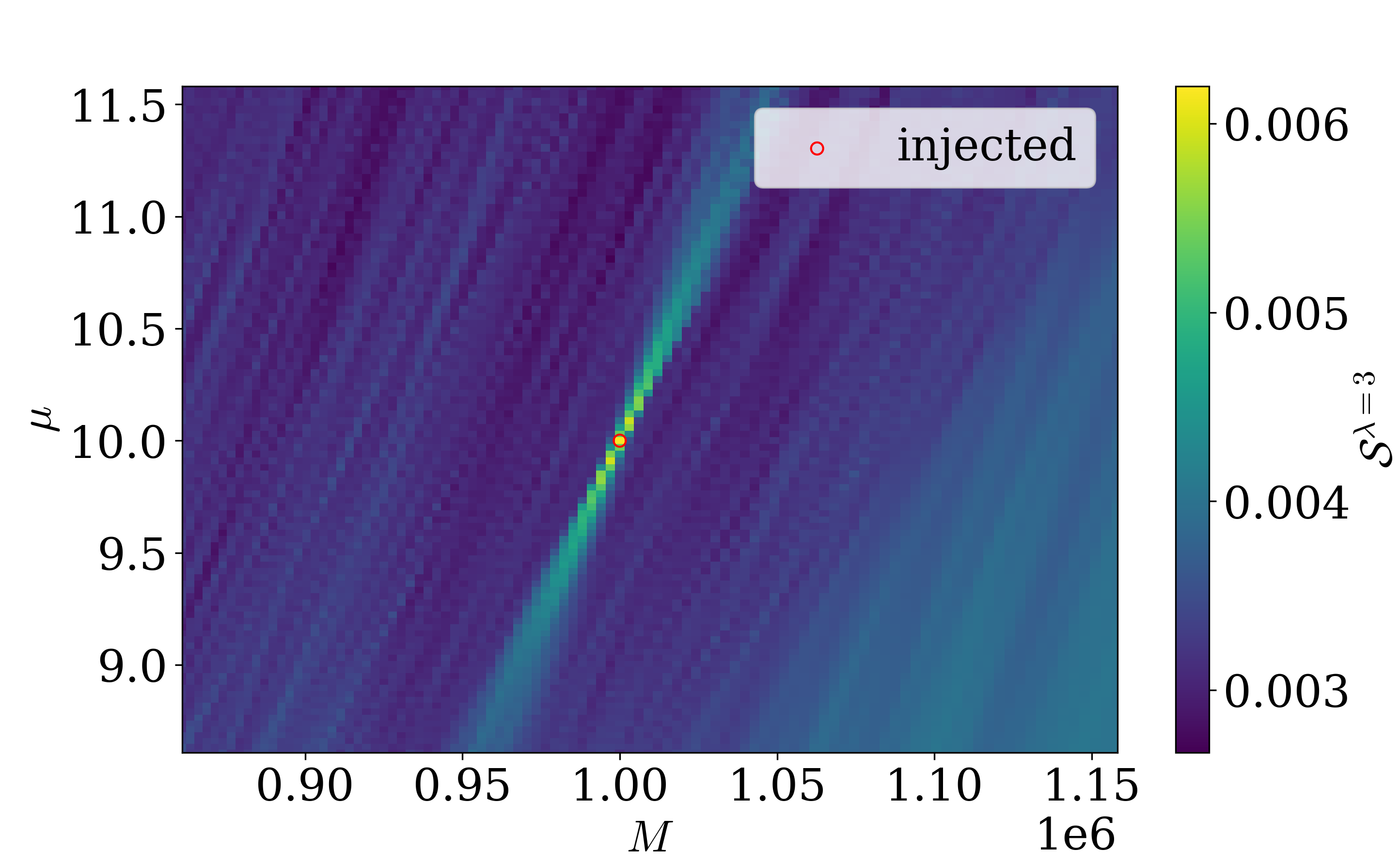}}

    \subfloat{\includegraphics[width=0.5\textwidth]{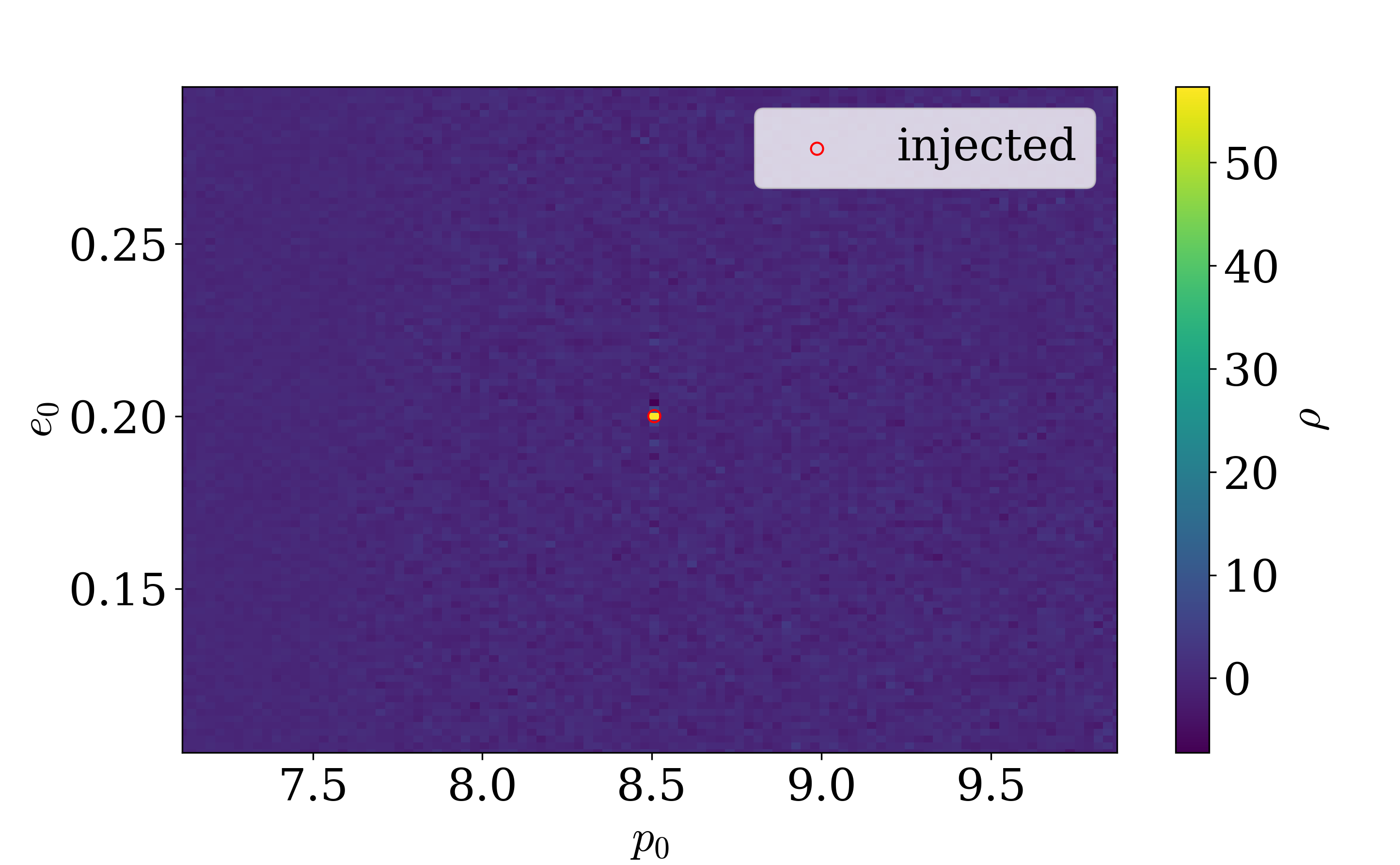}}
    \subfloat{\includegraphics[width=0.5\textwidth]{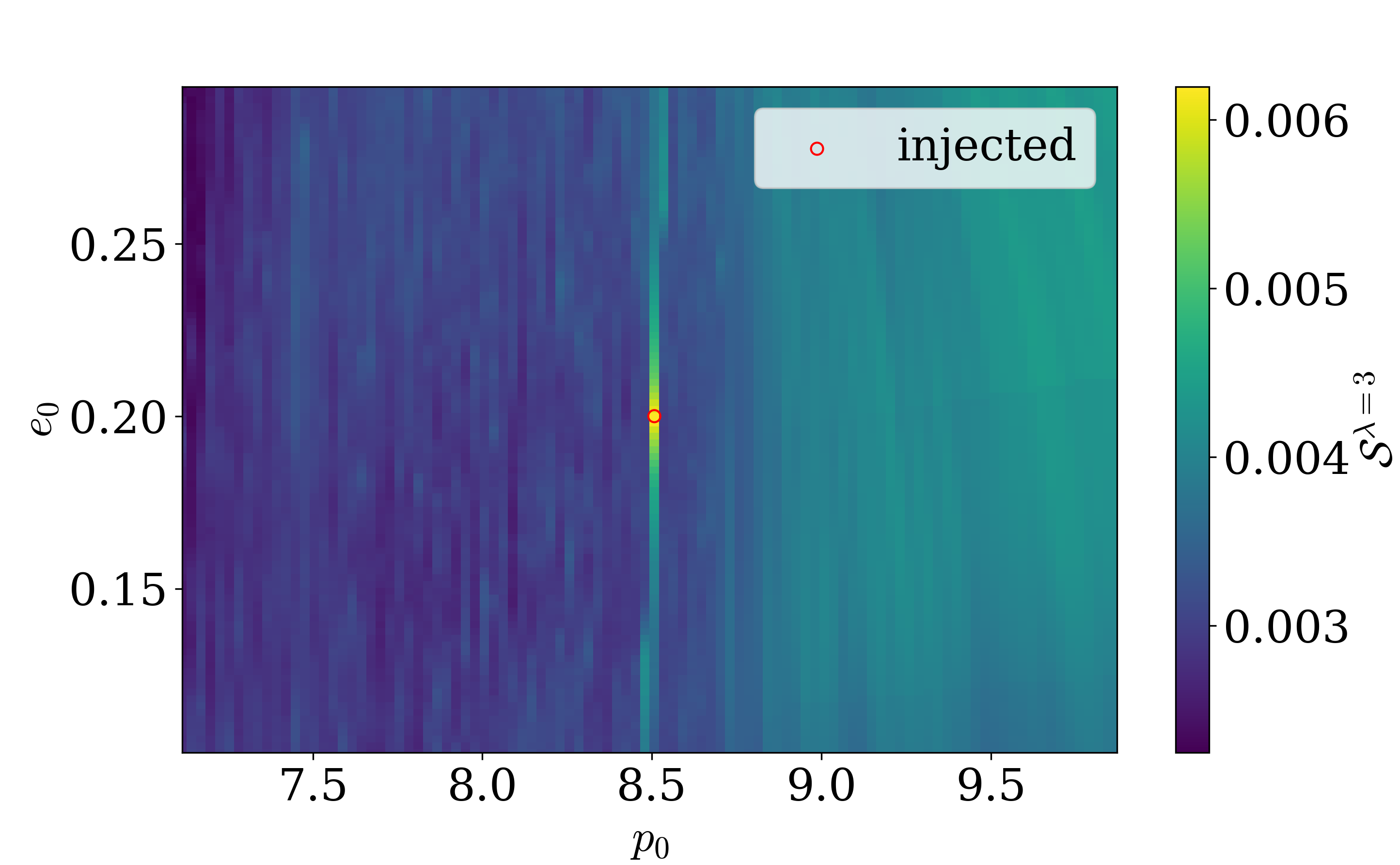}}


    \caption{Search functions computed varying only two parameters at a time. The left panels show the SNR function, while the right panels display the power spectrum matching. The top panels vary the masses, while the middle and bottom panels vary the eccentricity and semi-latus rectum.
    }
    \label{fig:function}
\end{figure*}

The difficulty in extracting extreme mass-ratio inspirals arises from the nonlocal parameter degeneracies and the fact that both the primary and secondary modes for standard search statistics are extremely small compared to the search volume. The narrow primary peak of the SNR statistic makes it particularly difficult to find the primary. Our goal is to increase the size of the primary peak by changing the search statistic.

Instead of using the fully coherent search statistic $\rho$ of Eq. \ref{eq:SNR}, we propose to maximize the following new search statistic
\begin{equation}
\label{eq:S}
\mathcal{S}^\lambda \coloneqq   \frac{ \langle  d ,  s(\theta) \rangle_\mathrm{tf,\lambda} } 
{\sqrt{\langle  s(\theta) ,  s(\theta) \rangle_\mathrm{tf,\lambda}}} \, ,
\end{equation}

with the time-frequency inner product defined as
\begin{equation}\label{eq:tfinner}
\langle  x ,  y \rangle_\mathrm{tf,\lambda}  \coloneqq \Bigg [\sum_{\tau=0}^{N-1}  4 \mathcal{R} \int_0^\infty 
\Bigg (
\frac{|\tilde x_\tau(f)| |\tilde y_\tau(f)|}{S_n(f)}
\Bigg )^\lambda
\, df 
\Bigg ]^{1/\lambda} \, ,
\end{equation}
with $\tilde x_\tau(f)$ being the short Fourier transform of the time series $x(t)$ for the first time window $\tau$.

We adopt the short-time Fourier transform representation of the data in order to find the best parameters that match the power in each time-frequency ``pixel''. Therefore, the function $\mathcal{S}$ has the advantage of disregarding the phase coherence between pixels but still imposes the frequency evolution coherence. This provides a wider primary mode in the search landscape for parameters $\{M, \mu, p_0, e_0\}$ as shown in Fig.~\ref{fig:function} for the source of Table~\ref{tab:injection}. 

We observe that the primary peak for the SNR function, $\rho$, is narrow compared to the search space. Instead, for $\mathcal{S}$, the primary mode is significantly wider, increasing the chance of convergence for a stochastic search.

The function $\mathcal{S}$ can be used to optimize parameters $\{M, \mu, p_0, e_0\}$, allowing us to refine the prior distribution before conducting a follow-up search optimizing the SNR $\rho$.

\begin{figure}[!ht]
    \centering
    \includegraphics[width=0.5\textwidth]{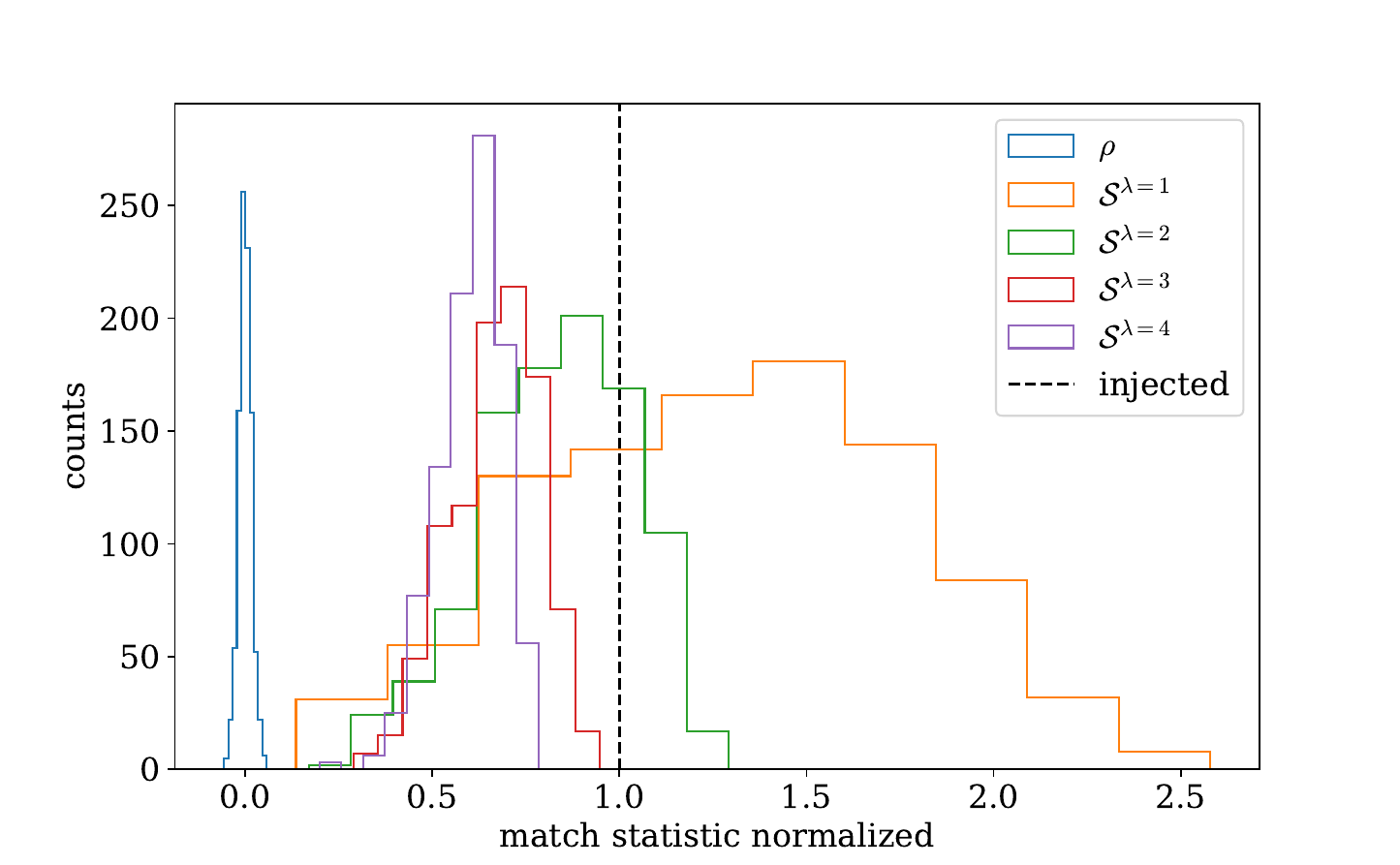}
    \caption{Histogram of different detection statistics for 1000 randomly drawn samples from the prior distribution. Each statistic is normalized with respect to its value at injection.}
    \label{fig:histogram}
\end{figure}

The exponent $\lambda$ acts as a generalized mean \cite{de_carvalho_2016_895400,Bullen2003} and can be adjusted to enhance the efficiency of the parameter optimization algorithm. As illustrated in Fig.~\ref{fig:histogram}, when drawing 1000 random samples from the prior, $\mathcal{S}^{\lambda=1}$ and $\mathcal{S}^{\lambda=2}$ sometimes produce higher match statistics for incorrect parameter sets than for the injected values. This suggests that these values of $\lambda$ are unsuitable, as they allow secondary modes or noise realizations to dominate over the true parameters. Increasing \( \lambda \) mitigates this effect by suppressing noise-induced matches, effectively raising the threshold for acceptable match accuracy. As an additional effect, for larger $\lambda$, the primary mode becomes narrower, making it more challenging to locate during optimization. As seen in the bottom left panel of Fig.~\ref{fig:function}, $\mathcal{S}^{\lambda=1}$ exhibits secondary modes that exceed the match statistic of the injected parameters. Meanwhile, the bottom right panel shows that $\mathcal{S}^{\lambda=4}$ results in a narrower primary mode compared to $\mathcal{S}^{\lambda=3}$.

To balance these effects, we set $\lambda=3$ for the remainder of this study, as it is the smallest value for which secondary modes do not exceed the match statistic of the injected parameters.

A limitation of optimizing \( \mathcal{S} \) is that it inherently reduces sensitivity to certain signals. Specifically, it restricts detection to those with sufficiently high SNR, potentially missing weaker or more diffuse signals. An extensive investigation of this search statistic through the entire EMRI parameter space is left for future work.

\subsection{Noise Estimation}
\label{sec:noise}

The noise estimation process is performed across the full frequency band, following a modified approach based on \cite{strub2024global}, originally introduced in \cite{strub2023accelerating}. In the following, we provide a concise overview of the pipeline and the modifications applied.

The noise estimation process begins with Welch’s method, where the initial power spectral density (PSD) is computed. To mitigate outlier peaks, we employ a 30-bin moving window, replacing extreme values with the median while iteratively shifting by 15 bins. To further smooth the estimate, a Savitzky-Golay filter is applied. The final residual noise curve is then obtained via spline interpolation.

Unlike the approach in \cite{strub2024global}, we modify Welch’s method by replacing the fixed window size of 15,000 samples with a dynamic size of $75{,}000 / dt$, where $dt$ is the data cadence. This allows for varying signal sampling rates.

Additionally, to further refine the estimated noise, we apply a second Savitzky-Golay filter after the first. This secondary filter has a polynomial degree of 1 and a window length of 5, improving the smoothness of the final noise curve.

\begin{figure}[!ht]
    \centering
    \includegraphics[width=0.5\textwidth]{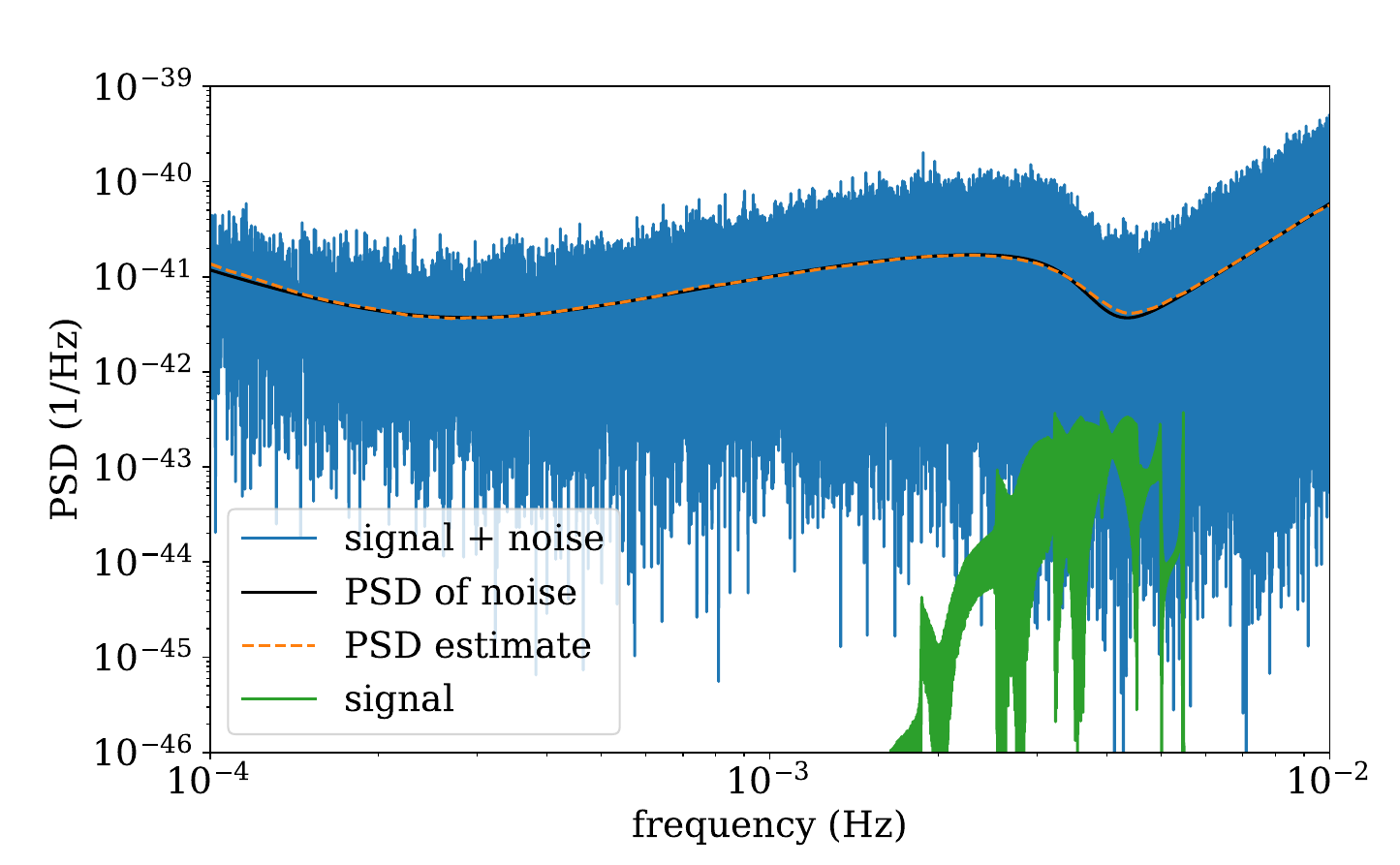}
    \caption{Noise estimation for the data shown in Fig.~\ref{fig:data}. The data (noise and signal realizations) are shown in blue. The assumed and recovered PSD are shown in black and dashed orange, respectively, for the TDI A channel. In green we show the signal of the EMRI.
    }
    \label{fig:noise}
\end{figure}

Fig.~\ref {fig:noise} presents the noise estimate for the LDC2a data set. The estimated noise generally aligns with the PSD used to generate the noise realization with a relative error of $ 3\%$ within the frequency range of $ \left[ \SI{0.4}{mHz}, \SI{10}{mHz} \right]$.

\subsection{The Pipeline}
\label{sec:pipeline}

To search for a signal, we need to define some priors. Here, we adopt the one shown in 
in Table \ref{tab:prior}.  Most are linearly uniform, with exceptions: $\theta_S$ and $\theta_K$ follow a cosine-uniform distribution, while $M$ and $\mu$ are log-uniform. 
Furthermore, the prior boundaries for the initial eccentricity $e_0$ are limited to the validity region of the current waveform simulator. 


Initially, the noise power spectral density $S_n(f)$ is estimated, which is a key component for calculating the scalar products in Eqs.~\ref{eq:scalar} and \ref{eq:tfinner}. For the analysis presented here, we first optimize $\mathcal{S}_\textrm{noresponse}$, where we compute $\mathcal{S}$ directly using the simulated GW signal, disregarding the LISA response. This approach restricts the search for parameters to the intrinsic variables $\{M, \mu, p_0, e_0\}$, bypassing the need for sky location matching. 

Sampling over \( p_0 \) within the full prior range can lead to signals that fall outside the valid domain of \texttt{few}. To address this, the first iteration of the search restricts the prior on the time to plunge, \( t_p \), to a two-week window. The value of \( p_0 \) is then determined numerically from the set \( \{t_p, M, \mu, e_0\} \) using Brent's method and by evolving the inspiral and identifying the value of \( p_0 \) that produces the specified \( t_p \). This function for calculating $p_0$ is implemented in \texttt{few}.


The initial search may become trapped in local maxima, far from the injected values. To mitigate this, we repeat the initial search optimization $\mathcal{S}_\textrm{noresponse}$ three times and choose the set of parameters with the highest value $\mathcal{S}_\textrm{noresponse}$. After initial optimization, we obtain estimates for the parameters $\{M, \mu, p_0, e_0\}$ that are close to the injected values. This allows us to narrow the prior and center it around the estimated parameters. From now on, we are sampling $p_0$ instead of $t_p$.

The next step involves optimizing $\mathcal{S}$, with the parameter set found previously serving as an initial guess. After refining the parameter space around the recovered values $\{M, \mu, p_0, e_0\}$, we perform a final optimization by maximizing the standard SNR $\rho$ as defined in Eq.~\ref{eq:SNR}. The optimization of $\mathcal{S}_\textrm{noresponse}$ and $\mathcal{S}$ is relatively quick, which we limit to only 150 and 200 iterations, respectively, while optimizing $\rho$, we limit to 4000 iterations.

Once the optimization process is completed, the luminosity distance $D_L$ is determined according to Eq.~\ref{eq:maxD}. Finally, the parameter boundaries are further refined and an MCMC algorithm is used to obtain the posterior distribution. A detailed summary of the pipeline is presented in Algorithm~\ref{alg:EMRI} as pseudocode.

\begin{table}[!ht]
\caption{The prior distribution $\Theta_\text{prior}$ for EMRIs.}
\begin{ruledtabular}
\begin{tabular}{ccc}
           Parameter & Boundary &  \\ \hline
  $\log M$ ($\textup{M}_\odot$) & $[\log 5 \cdot 10^5$ ,   $\log 10^7]$\\
  $\mu$ ($\textup{M}_\odot$)&  $[\log 5, \log 100]$\\
  $p_0$ &  set by $\{t_p, M, \mu, e_0\}$\\
  $t_p$ &  $[0.42,0.46]$\\
  $e_0$ &  $[0.01, 0.5]$\\
  $D_L$ (Gpc) & $[0.01  ,  100]$ \\
    $\cos \theta_S$  &     $[-1  ,  1]$\\
  $\phi_S$&      $[0, 2 \pi]$\\
        $\Phi_{\varphi0}$&          $[0 , 2 \pi]$\\
        $\Phi_{r0}$&          $[0 , 2 \pi]$\\
    $\cos \theta_K$  &     $[-1  ,  1]$\\
  $\phi_K$&      $[0, 2 \pi]$\\\\
\end{tabular}
\label{tab:prior}
\end{ruledtabular}
\end{table}

\begin{algorithm*}[!ht]
\caption{The EMRI search algorithm to analyze a data segment $d$ with function $f$ to optimize. The $\argmax$ is found using the differential evolution algorithm.}\label{alg:EMRI}
$\textbf{Function} \ \textbf{\textit{EMRI\_search}}(d)$\\
Estimate the noise $S_n(f)$\\
$\theta_{\textrm{init}} $ start guess randomly drawn from prior\\
Set boundaries to prior \\
$\theta_\textrm{max} \gets  \argmax\limits_{\theta}  \mathcal{S^\mathrm{\lambda = 3}_\mathrm{no response}} (\theta, d) $ with start $\theta_{\textrm{init}}$ and choosing $p_0$ such that $t_p$ within a given two week window\\
Set boundaries of $\left\{M, \mu, p0, e0\right\}$ to $\left\{M (1\pm 1\%) , \mu (1\pm 10\%) , p_0\pm 0.1 , e_0\pm 0.05 \right\}$ of the found parameters\\
$\theta_\textrm{max} \gets  \argmax\limits_{\theta}  \mathcal{S^\mathrm{\lambda = 3}} (\theta, d) $ with start $\theta_\textrm{max}$\\
Set boundaries of $\left\{M, \mu, p0, e0\right\}$ to $\left\{M (1\pm 0.5\%) M, \mu (1\pm 5\%) , p_0\pm 0.05, e_0\pm 0.025 \right\}$ of the found parameters\\
$\theta_\textrm{MLE} \gets  \argmax\limits_{\theta}  \rho (\theta, d) $ with start $\theta_{\textrm{max}}$\\
Compute $D_L$ according to Eq. \eqref{eq:maxD} with $\theta_\textrm{MLE}$\\
Set prior boundaries of $\left\{M, \mu, p0, e0, D_L\right\}$ to $\left\{M (1\pm 0.01\%), \mu (1\pm 0.1\%), p_0\pm 0.001, e_0\pm 0.001, D_L (1\pm 10\%) \right\}$ of the found parameters\\
Use the found $\theta_\textrm{MLE}$ as the initial proposal for an MCMC algorithm within the reduced boundaries to obtain the posterior\\
$\textbf{return } \theta_\textrm{MLE}$ and MCMC chain
\end{algorithm*}

After successfully detecting a signal, we perform an MCMC analysis. For this purpose, we use the $\texttt{Eryn}$ MCMC sampler \cite{karnesis2023eryn, michael_katz_2023_7705496, foreman2013emcee}, configured with four temperatures, 32 walkers, 3000 samples, and 1000 burn-in samples. The resulting chain is used to compute the posterior distribution.


\section{Results}  
\label{sec:results}  

For this analysis, we adopted a sampling interval of $dt = \SI{50}{s}$, under the assumption that no significant signal components exceed the Nyquist frequency, $f_{\text{Nyquist}} = 1/(2dt) = \SI{10}{mHz}$. Time-frequency analysis was performed using the \texttt{stft} function from the \texttt{SciPy} library \cite{virtanen_scipy_2020}. We used a Hann window with an overlap of $50\%$ and set the segment length to $50000 / dt$, optimizing the resolution in both the time and frequency domains.

The boundary of the time to plunge $t_p$ is set to be a two week interval. Either the pipeline for obtaining $\theta_\textrm{MLE}$ is repeatedly performed with a moving window for the $t_p$ prior, or a previous detection algorithm determined an EMRI signal within a given time interval for $t_p$. If no previous detection has been made, the algorithm can still be applied, and if the identified signal has an SNR above a certain threshold the signal is determined as a detection. In this article, we are using differential evolution (DE) \cite{storn1997differential} to maximize $\mathcal{S}_\mathrm{no response}$, $\mathcal{S}$, and $\rho$.

\subsection{Extraction of an SNR 56 Signal}  

\begin{figure*}[!ht]
    \includegraphics[width=1\textwidth]{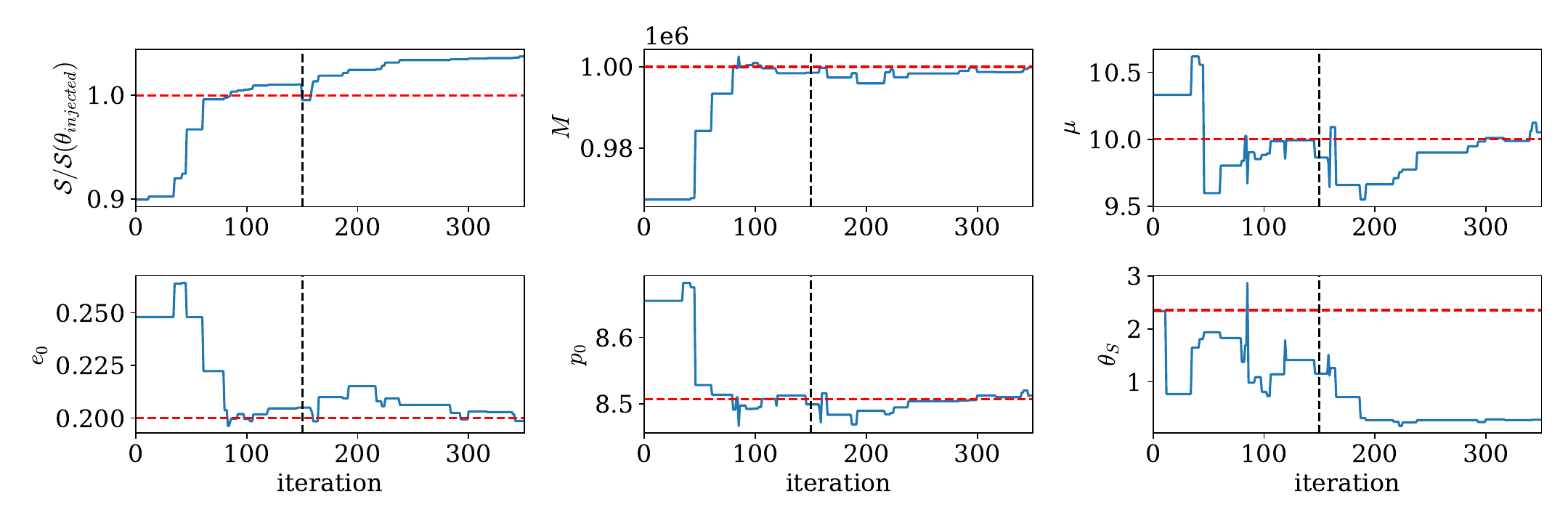}
\caption{The search chain of optimizing $\mathcal{S}_\textrm{noresponse}$ in the first 150 iteration and $\mathcal{S}$ from iteration 151 to 350. The black vertical lines show the change in the optimization function. The dashed red lines show the injected values. The top left panel shows the optimization metric normalized by the value of the injection.
}
\label{fig:chainSall}
\end{figure*}

\begin{figure*}[!ht]
    \includegraphics[width=1\textwidth]{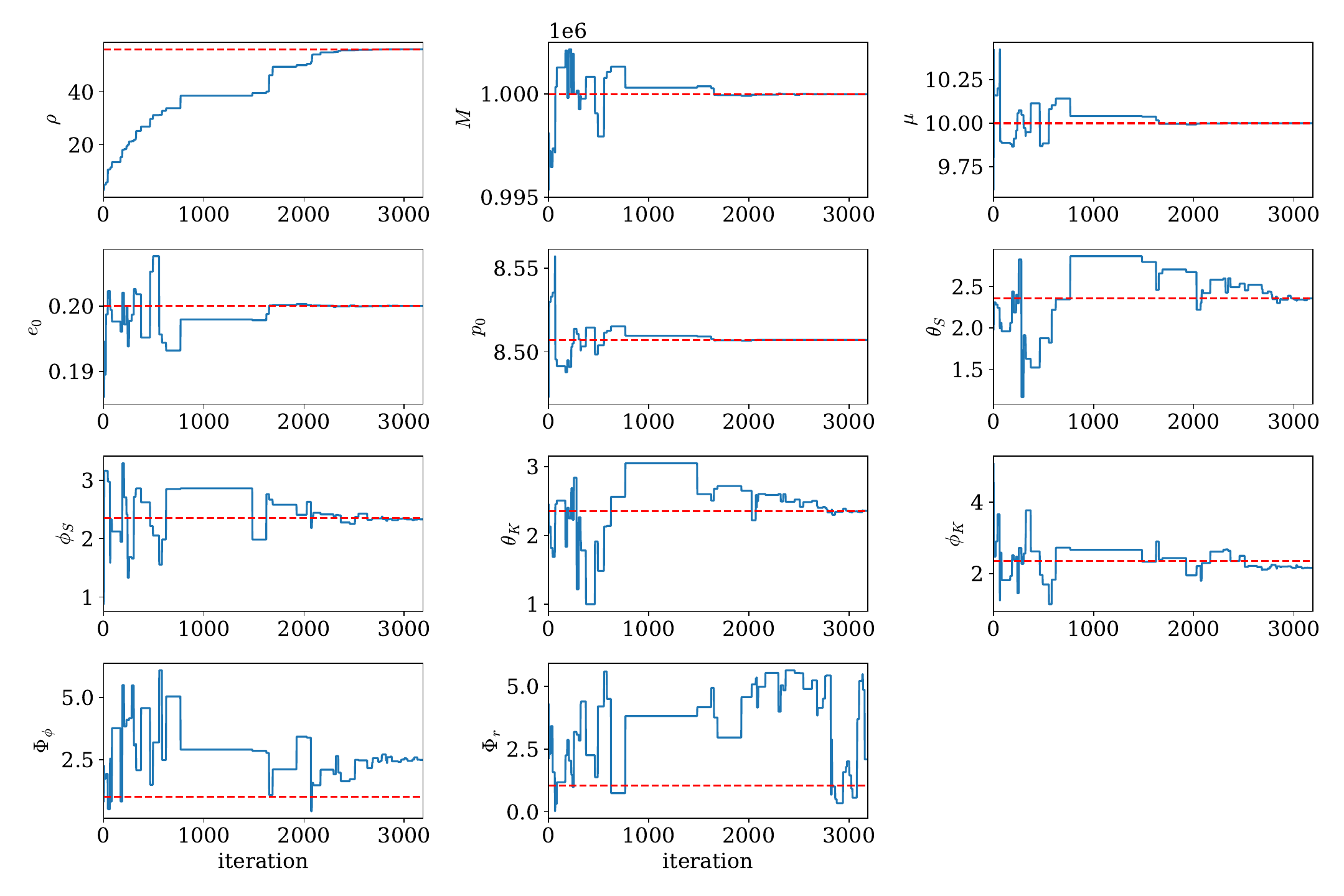}
\caption{The search chain of optimizing the SNR $\rho$. The dashed red lines show the injected values.}
\label{fig:chain_snr}
\end{figure*}

To assess signal recovery, we applied our detection pipeline to the dataset illustrated in Fig. \ref{fig:data}, using injection parameters corresponding to Table \ref{tab:injection}. The iterative search for parameters, visualized in Fig. \ref{fig:chainSall} and Fig. \ref{fig:chain_snr}, enabled an iterative reduction in the prior distribution of key parameters $\left\{M, \mu, p_0, e_0 \right\}$. This refinement helped identify the primary mode of the SNR function, $\rho$, and ultimately led to the convergence of the maximum likelihood estimate $\theta_{\text{MLE}}$, with recovered parameters closely matching the injected values.

\begin{figure}[!ht]
    \includegraphics[width=0.5\textwidth]{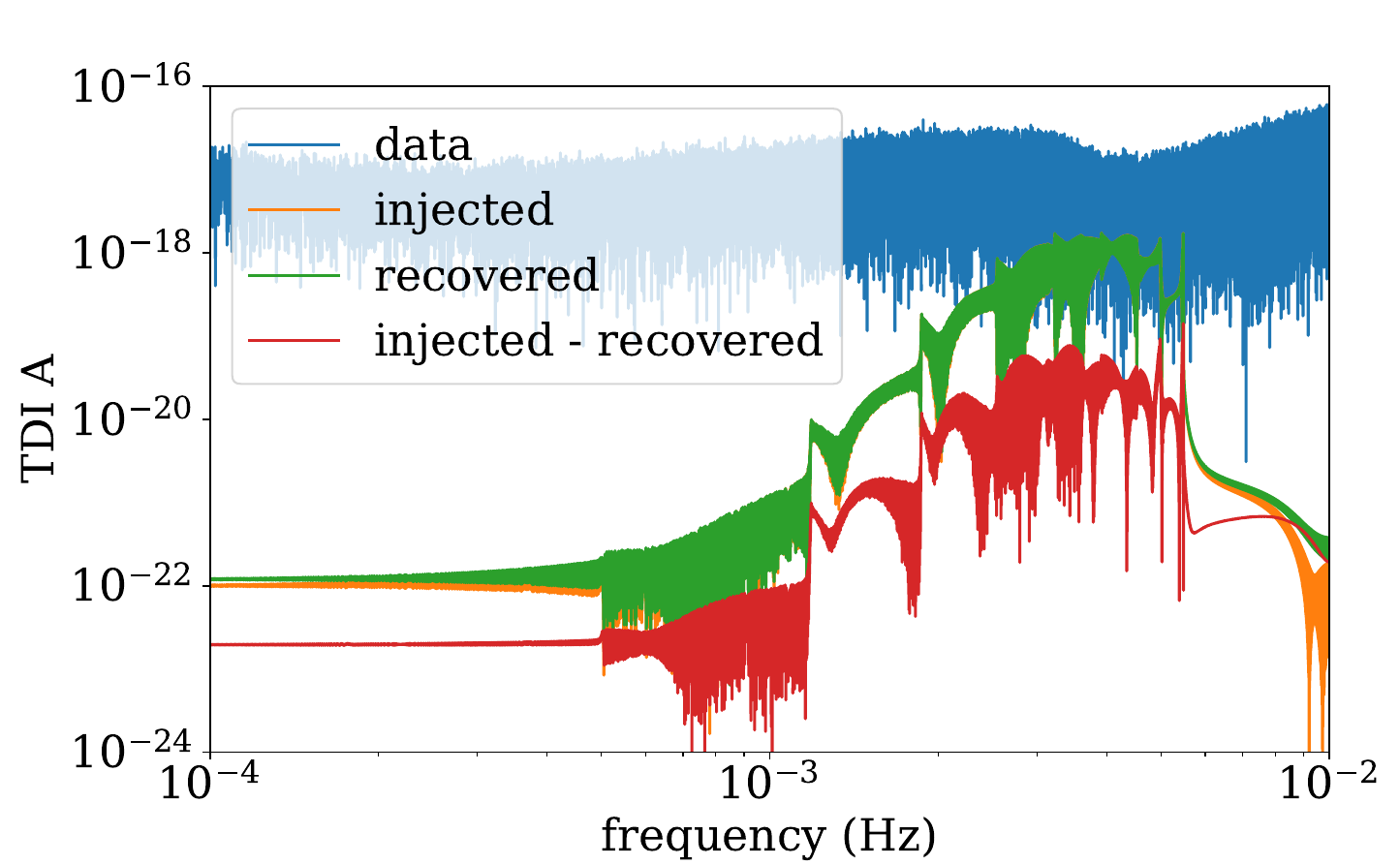}
\caption{The data, injected and recovered signal in the frequency domain. The signal parameters are listed in Table \ref{tab:found}.}
\label{fig:loglogf}
\end{figure}


\begin{table}[!ht]
\caption{The injected and recovered parameters of the EMRI signal. The uncertainties are computed with the MCMC chain, which is shown as a corner plot in Fig. \ref{fig:corner}. Here we show the detected SNR $\rho$ according to Eq. \ref{eq:SNR}.}
\begin{ruledtabular}
\begin{tabular}{cccc}
           Parameter &  Injected & Recovered  \\ \hline
  $M$ ($\textup{M}_\odot$) & $10^6$ & 999991 $\pm$ 20 \\
  $\mu$ ($\textup{M}_\odot$) &  $10$ & 9.9998  $\pm$ 0.0011\\
  $e_0$ &  $0.2$ & 0.20003  $\pm$ 0.000075\\
  $p_{0}$ &  $8.5072$ & 8.50719  $\pm$ 0.00006\\
  $D_L$ (Gpc) & 0.85  & 0.855 $\pm$ 0.02\\
    $ \cos \theta_S$  &     -0.71 & -0.71  $\pm$ 0.06\\
  $\phi_S$&      2.36 & 2.33  $\pm$ 0.07 \\
        $\Phi_{\varphi0}$&          $1$ & 2.5  $\pm$ 0.7\\
        $\Phi_{r0}$&          $3$ & 3.0 $\pm$ 0.3\\
    $\cos \theta_K$  &     -0.71 & -0.71 $\pm$ 0.09\\
  $\phi_K$&      2.36 & 2.17 $\pm$ 0.09\\
  $\rho$& $56.09$ & 56.12\\
\end{tabular}
\label{tab:found}
\end{ruledtabular}
\end{table}

The recovered parameters are presented in Table \ref{tab:found} and visualized in the frequency domain in Fig. \ref{fig:loglogf}. The injected and recovered parameters are in close agreement, where the difference is one order of magnitude lower than the injected signal. Notably, the detected SNR of the recovered signal slightly surpasses that of the injected signal, which can be attributed to noise fluctuations. The condition $\rho(\theta_{\textrm{MLE}}) > \rho(\theta_{\textrm{injected}})$ further validates the success of the optimization procedure in refining the parameter search.

\begin{figure}[!ht]
    \includegraphics[width=0.5\textwidth]{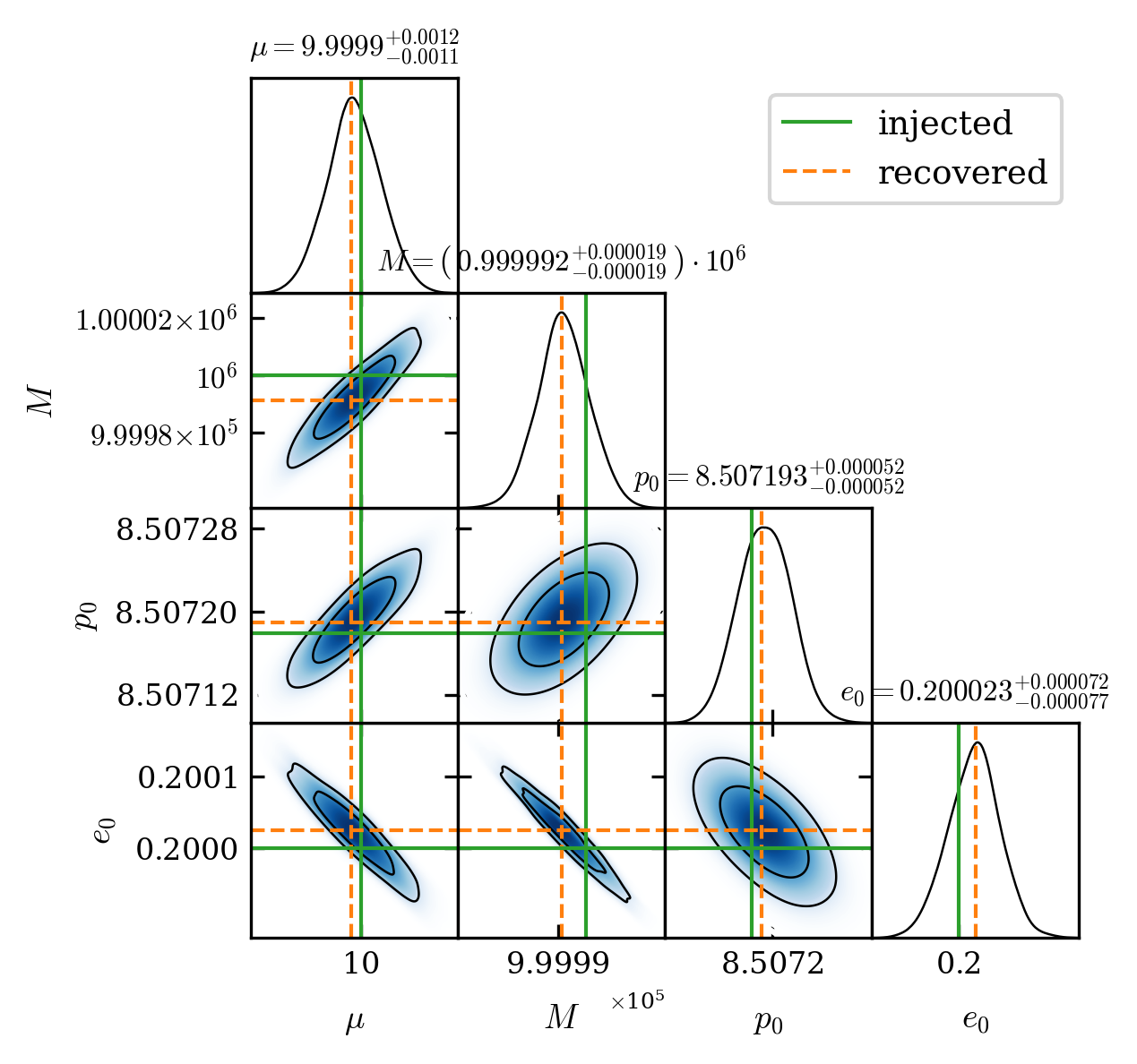}
\caption{Green solid lines denote the injected parameters, while orange dashed lines correspond to the recovered parameters, $\theta_{\textrm{MLE}}$, presented in Table~\ref{tab:found}. The blue contours illustrate the $68\%$ and $95\%$ confidence intervals for the parameter estimates. The posterior distribution is computed using $\theta_{\textrm{MLE}}$ as the initial condition for the MCMC sampling. The full posterior Figure \ref{fig:cornerfull} with all parameters can be found in the appendix \ref{appendix:posteriors}.
}
\label{fig:corner}
\end{figure}

To finalize the results with a posterior distribution, we employed $\texttt{Eryn}$ to generate an MCMC chain. The resulting corner plot in Fig. \ref{fig:corner} excludes some parameters for better visualization. The complete corner plot with all parameters is added to Appendix \ref{appendix:posteriors} shown in Fig. \ref{fig:cornerfull}. It is noticeable that the injected parameters are within the $68\%$ probability region of the parameter estimation for all parameters except the phase $\Phi_{\varphi0}$.

\subsection{Increasing the mass}

\begin{figure*}[!ht]
    \includegraphics[width=1\textwidth]{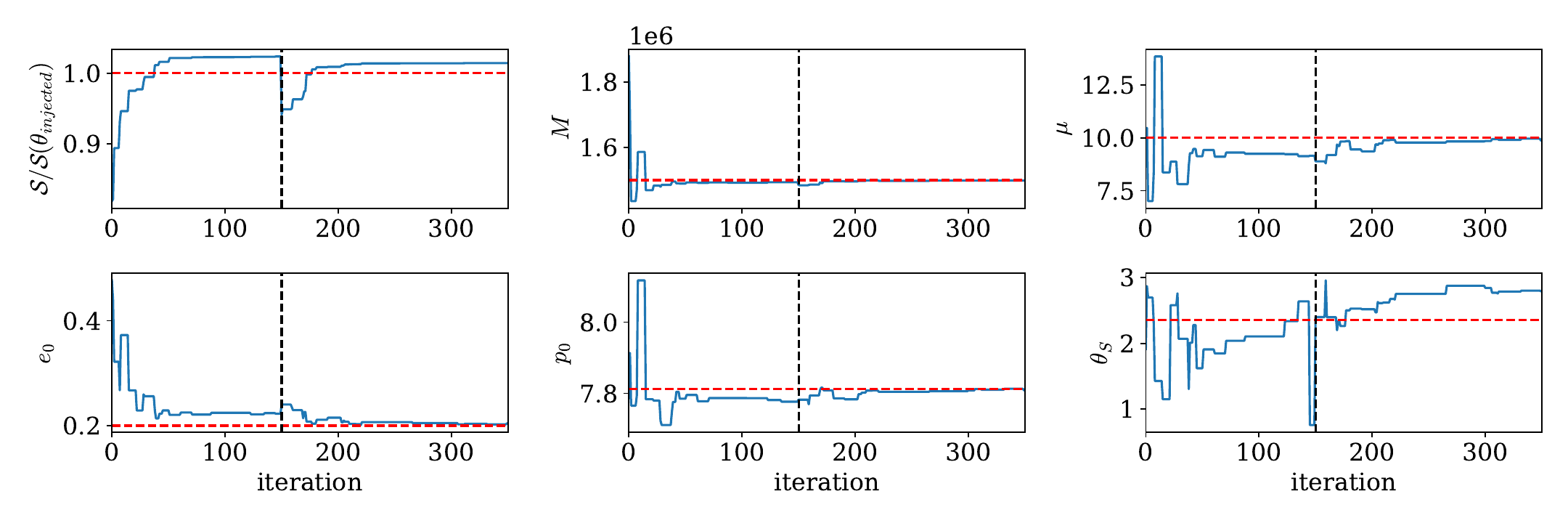}
\caption{Search for the increased mass signal. The resulting chain optimizes $\mathcal{S}_\textrm{noresponse}$ during the first 150 iterations, followed by optimization of $\mathcal{S}$ from iteration 151 to 350. Black vertical lines indicate the transition between the two objective functions. Dashed red lines mark the injected parameter values. The top-left panel displays the optimization metric, normalized by its value at the injection.
}
\label{fig:chainSM}
\end{figure*}

\begin{figure*}[!ht]
    \includegraphics[width=1\textwidth]{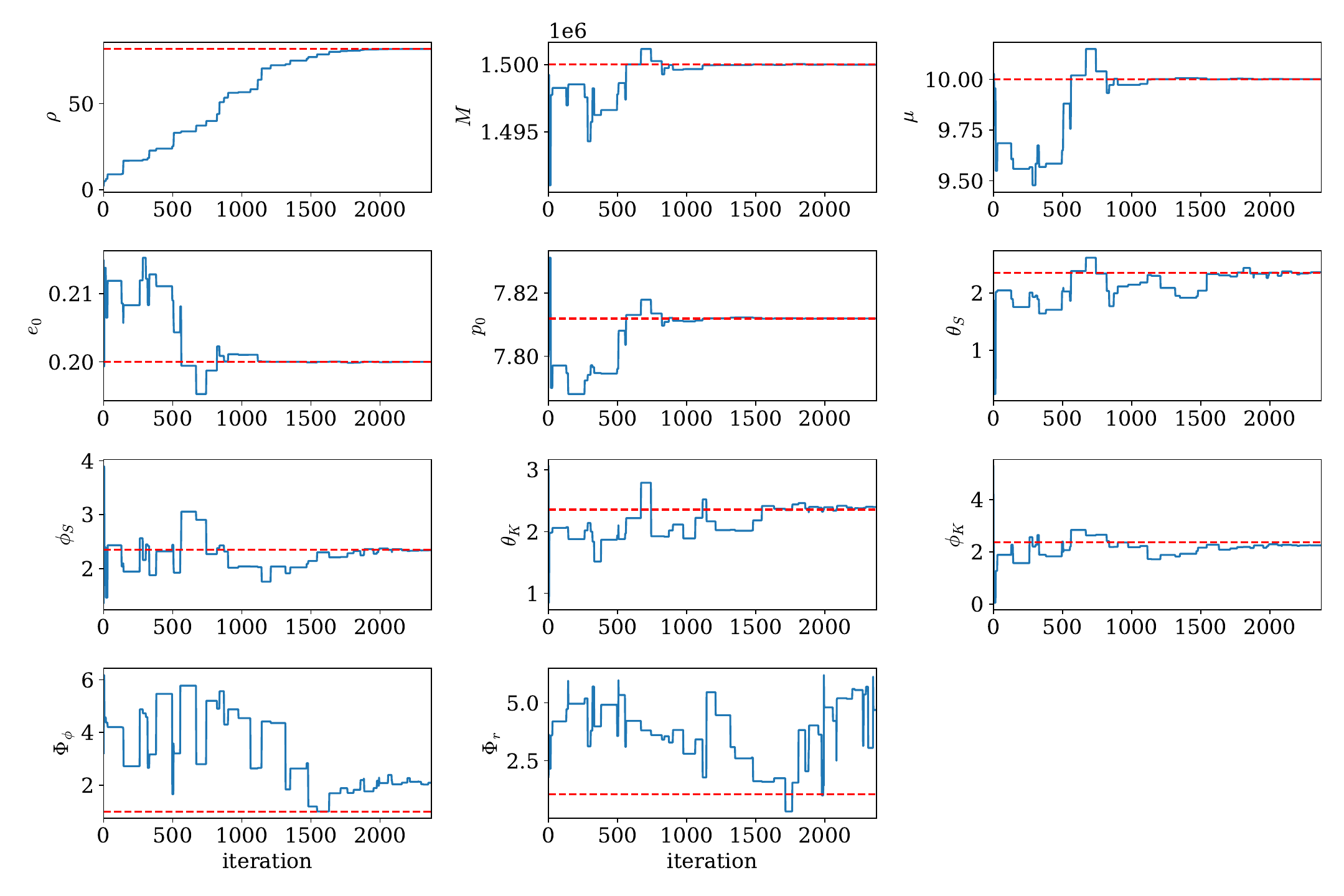}
\caption{Search for the increased mass signal, performed by optimizing the SNR $\rho$. The dashed red lines indicate the injected parameter values.}
\label{fig:chainSNRM}
\end{figure*}

In a second example, we modify the EMRI by increasing the mass of the massive black hole and setting the distance to $\SI{0.2}{Gpc}$. The differential evolution algorithm optimizing $\rho$ completes after 2,373 iterations, resulting in a reduced computational time. The injected and recovered parameters are summarized in Table \ref{tab:foundHigherM}. Figures \ref{fig:chainSM} and \ref{fig:chainSNRM} illustrate the search chains leading to the recovered values. We then calculate the posterior distribution using an MCMC chain, with the corner plot shown in Fig. \ref{fig:cornerM} in the Appendix \ref{appendix:posteriors}. Notably, the injected parameters fall within the $68\%$ confidence interval for all parameters except for the phase $\Phi_{\varphi0}$. The accuracy and precision of the recovered parameters are higher than those of the first signal, probably because of the higher SNR of the injected signal.

\begin{table}[!ht]
\caption{The injected and recovered parameters of the EMRI signal. The uncertainties are computed with the MCMC chain, which is shown as a corner plot in Fig. \ref{fig:corner}.}
\begin{ruledtabular}
\begin{tabular}{cccc}
           Parameter &  Injected & Recovered  \\ \hline
  $M$ ($\textup{M}_\odot$) & $1500000$ & 1499998 $\pm$ 15 \\
  $\mu$ ($\textup{M}_\odot$) &  $10$ & 10.0003  $\pm$ 0.0009\\
  $e_0$ &  $0.2$ & 0.2  $\pm$ 0.00005\\
  $p_{0}$ &  $7.8119$ & 7.8119  $\pm$ 0.00003\\
  $D_L$ (Gpc) & 0.2  & 0.201 $\pm$ 0.006\\
    $ \cos \theta_S$  &     -0.71 & -0.71  $\pm$ 0.03\\
  $\phi_S$&      2.36 & 2.34  $\pm$ 0.04 \\
        $\Phi_{\varphi0}$&          $1$ & 2.1  $\pm$ 0.9\\
        $\Phi_{r0}$&          $3$ & 2.9 $\pm$ 0.2\\
    $\cos \theta_K$  &     -0.71 & -0.74 $\pm$ 0.05\\
  $\phi_K$&      2.36 & 2.24 $\pm$ 0.12\\
  $\rho$& $81.62$ & 81.66\\
\end{tabular}
\label{tab:foundHigherM}
\end{ruledtabular}
\end{table}



\subsection{Computational cost}

The current setup requires 140 function evaluations per iteration for parameter optimization. Consequently, obtaining $\theta_\textrm{MLE}$ involves 510,000 function evaluations in 3,838 iterations. The time for the evaluation of each function is influenced by the efficiency of the waveform model, the LISA response computation, the statistic calculation of the match, and the computational hardware. With the implementation of $\texttt{few}$ and the use of a mobile computer with a Quadro RTX 4000 mobile GPU, the computation of $\rho$ and $\mathcal{S}$ takes approximately $\SI{67}{ms}$, while $\mathcal{S}_{\textrm{noresponse}}$ takes $\SI{34}{ms}$. Thus, optimizing $\mathcal{S}_{\textrm{noresponse}}$ for 450 iterations, $\mathcal{S}$ for 200 iterations, and $\rho$ for 3188 iterations takes roughly $\SI{10.9}{h}$ on a laptop. When searching for the signal with the parameters of Table \ref{tab:foundHigherM}, the optimization of $\rho$ concluded after 2,373 iterations, reducing the computational time to about $\SI{7.3}{h}$.

The time required to generate the posterior distribution is determined primarily by the desired chain length. For a chain with 3000 samples, 32 walkers, four temperatures, and 1000 burn-in samples, the process takes approximately $\SI{8}{h}$. Runtimes can be significantly reduced using high-performance computing resources.

\section{Conclusions}
\label{sec:conclusion}

The first successful identification and inference of EMRI signals was achieved by optimizing a novel match statistic, which helped alleviate the complexities arising from nonlocal parameter degeneracies. Our pipeline can detect and estimate the parameters of the EMRI signals, even in the presence of instrument noise and the galactic foreground. This accomplishment represents a significant step forward in the preparation for the LISA mission and opens up exciting prospects for probing the universe's most extreme environments. Detecting and analyzing EMRI signals will allow astrophysicists to test general relativity in strong gravitational fields, investigate the formation and evolution of supermassive black holes, and possibly uncover physics beyond the standard model of cosmology. Moreover, accurately identifying signals plays a crucial role in minimizing their interference by subtracting them from the data, facilitating the identification of other sources in a global analysis pipeline. This pipeline paves the way for studying the interplay between different types of gravitational wave sources, including EMRIs.

A current limitation of the method is its reliance on relatively high-SNR signals that stand out within a noisy power spectrum. In this study, we successfully estimated parameters for a signal with an optimal SNR of \( \rho_{\text{optimal}} = 56 \). Future work will aim to improve the sensitivity to weaker signals by refining the match statistic and exploring the performance limits of minimum SNR and signal duration. Additionally, incorporating spinning black holes into the analysis will introduce new complexities that will need to be addressed in future work. 

\section{Acknowledgements}
We are thankful for the GPU implementation of $\texttt{few}$ and all previous work on ERMI signal simulation and search, which enabled this work. This work was supported by GW-Learn, a project funded by a Sinergia grant from the Swiss National Science Foundation.


\appendix

\section{Posteriors}
\label{appendix:posteriors}
\begin{figure*}[!ht]
    \includegraphics[width=1\textwidth]{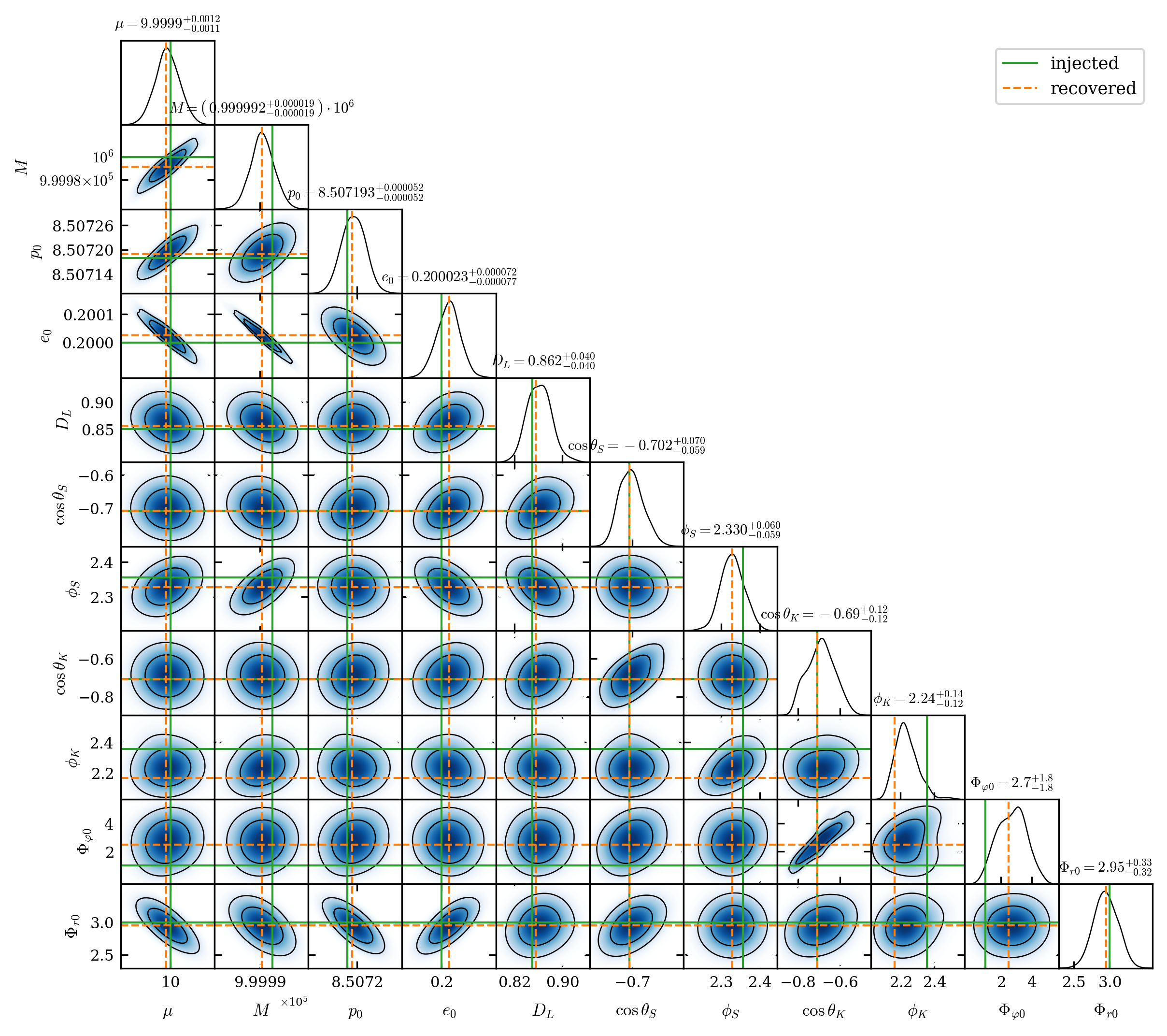}
\caption{The green solid lines indicate the injected parameters, while the orange dashed lines represent the recovered parameters, $\theta_{\textrm{MLE}}$, as listed in Table~\ref{tab:found}. The blue contours represent the $68\%$ and $95\%$ confidence regions of the parameter estimation. The posterior distribution is derived using $\theta_{\textrm{MLE}}$ as the starting point for the MCMC chain.
}
\label{fig:cornerfull}
\end{figure*}

\begin{figure*}[!ht]
    \includegraphics[width=1.1\textwidth]{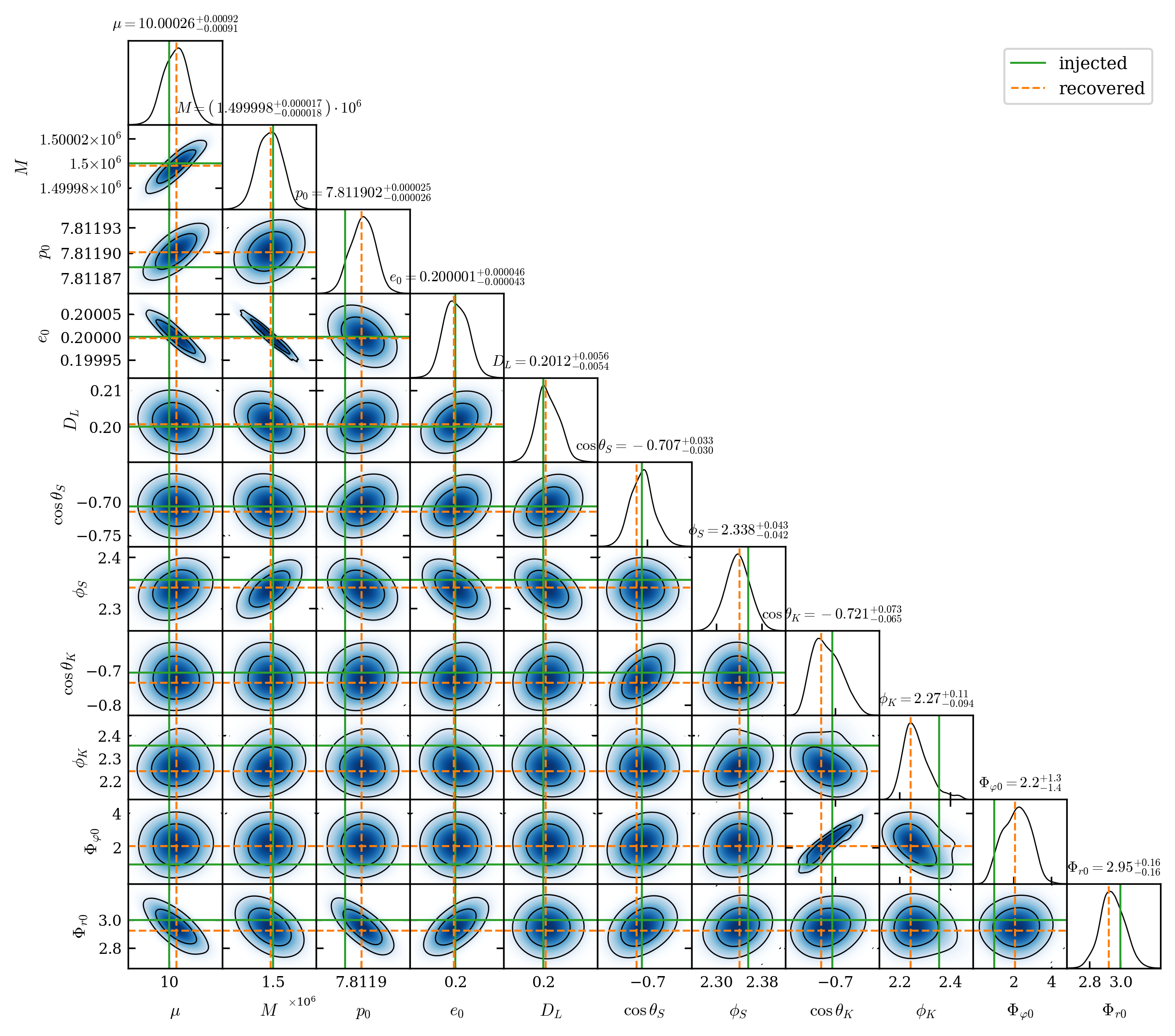}
\caption{Corner plot of the increased mass signal. The injected parameters are shown as green solid lines, while the recovered parameters, $\theta_{\textrm{MLE}}$, are indicated by orange dashed lines, where the values of the increased mass signal are presented in Table~\ref{tab:foundHigherM}. The blue contours depict the $68\%$ and $95\%$ confidence regions of the parameter estimation. The posterior distribution is generated using $\theta_{\textrm{MLE}}$ as the initial point for the MCMC chain.}
\label{fig:cornerM}
\end{figure*}

We present the full posterior distributions of the samples discussed in the results. Figure \ref{fig:cornerfull} shows the posterior distributions for all signal parameters corresponding to the injection values listed in Table \ref{tab:injection}. Similarly, Figure \ref{fig:cornerM} displays the posterior distributions for the signal parameters associated with the values in Table \ref{tab:foundHigherM}.

\bibliography{references}

\end{document}